\documentclass{aa}
\usepackage{rotating}
\usepackage{graphicx}
\usepackage{natbib}
\bibpunct{(}{)}{;}{a}{}{,}
\begin{document}
\titlerunning{AXIS medium survey}
   \title{The XMM-Newton Serendipitous Survey }

    \subtitle{II. First results from the AXIS high galactic latitude medium
    sensitivity survey\thanks{Based on observations obtained with
    XMM-Newton, an ESA science mission with instruments and
    contributions directly funded by ESA Member States and the USA
    (NASA)}}

\authorrunning{X. Barcons et al}
   \author{X. Barcons		\inst{1}
	\and
	 F.J. Carrera		\inst{1} 
	\and
	 M.G. Watson		\inst{2}
	\and
	 R.G. McMahon		\inst{3}
	\and
	 B. Aschenbach		\inst{4}
	\and
	 M.J. Freyberg		\inst{4}
	\and
	 K. Page		\inst{2}
	\and
         M.J. Page		\inst{5}
	\and
	 T.P. Roberts		\inst{2}
	\and
         M.J.L. Turner		\inst{2}
	\and
	D. Barret		\inst{6}
	\and
	H. Brunner		\inst{7}
	\and
         M.T. Ceballos		\inst{1}  
	\and
	 R. Della Ceca 		\inst{8}
	\and 
	P. Guillout		\inst{9}
	\and
	G. Hasinger 		\inst{4}, \inst{7} 
	\and 
	T. Maccacaro		\inst{8}
	\and
	S. Mateos		\inst{1}
	\and
	C. Motch         	\inst{9}  
	\and 
	I. Negueruela		\inst{9}
	\and
	 J.P. Osborne		\inst{2}
	\and
	 I. P\'erez-Fournon	\inst{10} 
	\and
	A. Schwope 		\inst{7}
	\and
        P. Severgnini 		\inst{11}
	\and 
	G.P. Szokoly 		\inst{7}
	\and
	N.A. Webb		\inst{6}
	\and
	P.J. Wheatley 		\inst{2}
	\and
	D.M. Worrall		\inst{12}
	}

   \offprints{X. Barcons,
	\email{barcons@ifca.unican.es}}

   \institute{Instituto de F\'\i sica de Cantabria (CSIC-UC), 39005 Santander, Spain  
        \and Department of Physics and Astronomy, University of Leicester, LE1 7RH, UK 
	\and Institute of Astronomy, Madingley Road, Cambridge CB3 0HA, UK 
	\and Max-Planck Institut f\"ur Extraterrestriche Physik,
        Postfach 1312, 85741 Garching, Germany 
	\and Mullard Space Science Laboratory, UCL, Holmbury St Mary,
        Dorking, Surrey RH5 6NT, UK 
	\and Centre d'Etude Spatiale des Rayonnements, 9 Avenue
        du Colonel Roche, 31028 Toulouse Cedex 04, France
	\and Astrophysikalishes Institut Potsdam, An der Sternwarte
        16, 14482 Potsdam, Germany 
	\and Osservatorio Astronomico di Brera, via Brera 28, 20121 Milano, Italy 
	\and Observatoire Astronomique de Strasbourg, 11 rue de
        l'Universit\'e, 67000 Strasbourg, France 
	\and Instituto de Astrof\'\i sica de Canarias, 38200 La Laguna, Tenerife, Spain 
	\and Dipartimento di Astronomia e Scienza dello Spazio,
        Universit\`a di Firenze, Largo E. Fermi 5,
        50125 Firenze, Italy 
	\and Department of Physics, University of Bristol, Royal Fort,
        Tyndall Avenue, Bristol, BS8 1TL, UK
             }

   \date{2 November 2001}

   \abstract{ We present the first results on the identifications of a
medium sensitivity survey (X-ray flux limit $2\times 10^{-14}\, {\rm
erg}\, {\rm cm}^{-2}\, {\rm s}^{-1}$ in the 0.5-4.5 keV band) at high
galactic latitude ($\mid b\mid>20^{\circ}$) carried out with the
XMM-Newton X-ray observatory within the AXIS observing programme. This
study is being conducted as part of the XMM-Newton Survey Science
Centre activities towards the identification of the sources in the
X-ray serendipitous sky survey. The sample contains 29 X-ray sources
in a solid angle of $0.26\deg^2$ (source density $113\pm 21$
sources~$\deg^{-2}$), out of which 27 (93\%) have been identified.
The majority of the sources are broad-line AGN (19), followed by
narrow emission line X-ray emitting galaxies (6, all of which turn out
to be AGN), 1 nearby non-emission line galaxy (\object{NGC 4291}) and
1 active coronal star. Among the identified sources we find 2
broad-absorption line QSOs ($z\sim 1.8$ and $z\sim 1.9$), which
constitute $\sim$ 10\% of the AGN population at this flux level,
similar to optically selected samples. Identifications of a further 10
X-ray sources fainter than our survey limit are also presented.
 }

   \maketitle
%

\section{Introduction}

The XMM-Newton observatory, the second cornerstone of the Horizon 2000
science programme of the European Space Agency has been carrying out
science operations since early 2000.  Thanks to its high collecting
area, large field of view and
moderate angular and spectral resolution XMM-Newton is the most
powerful observatory in hard X-rays (photon energy $> 2$ keV), opening
an almost unexplored window to the Universe \citep{Jansen2001}. The
sensitivity to hard X-rays (not attained by previous missions like
$Einstein$ and $ROSAT$) allows the detection and study of the most
energetic objects in the Universe (Active Galactic Nuclei - AGN), most
of which are believed to be deeply hidden inside large amounts of
absorbing gas and inconspicuous at virtually all other wavelengths.

During science observations (with exposure time over $\sim
10$ ks)  with the EPIC cameras operating
in ``Full Frame'' mode \citep{Turner2001,Struder2001}
XMM-Newton is discovering $\sim 30-150$ new X-ray sources, which add
to the XMM-Newton serendipitous survey at an expected rate of $\sim
50000$ new sources per year. The XMM-Newton Survey Science Centre
(SSC) was appointed by ESA to exploit scientifically the XMM-Newton
serendipitous survey for the benefit of the scientific community and
as a major legacy of XMM-Newton to future generations.  This is being
tackled by the SSC consortium in terms of a mostly ground-based
optical follow-up and identification (XID) programme.

The XID programme has been described in detail in
\citet{Watson2001}. Briefly, its implementation has been divided into
two parts: a {\it core programme} which will identify --
spectroscopically -- significant samples of sources at X-ray flux
limits around $\sim 10^{-13}\, {\rm erg}\, {\rm cm}^{-2}\, {\rm s}^{-1}$
(bright sample), $\sim 10^{-14}\, {\rm erg}\, {\rm cm}^{-2}\, {\rm
s}^{-1}$ (medium sample) and $\sim 10^{-15}\, {\rm erg}\, {\rm cm}^{-2}\,
{\rm s}^{-1}$ (faint sample) covering a range of galactic latitudes,
and an {\it imaging programme} aiming at providing deep
optical/infrared images in several colours of a large number of
XMM-Newton fields to facilitate statistical identifications of the
serendipitous sources.

AXIS (``An XMM-Newton International Survey")\footnote{See
http://www.ifca.unican.es/\~{}xray/AXIS} forms the backbone
of the XID programme by providing the ground-based resources that are
essential for the exploitation of the XMM-Newton serendipitous sky
survey.  Besides making a first and major contribution to the XID
programme, AXIS will define the quality standard and will guide future
steps in the implementation of the XID programme.  AXIS has been
conceived and designed to make optimal use of the available
instrumentation on the telescopes of the Observatorio del Roque de los
Muchachos. AXIS has been awarded a total of 85 observing nights spread
over the period April 2000 - April 2002 on the 4 larger telescopes of
the Observatorio del Roque de Los Muchachos: the 2.5m Isaac Newton
Telescope (INT), the 2.5m Nordic Optical Telescope (NOT), the 3.5m
Telescopio Nazionale Galileo (TNG) and the 4.2m William Herschel
Telescope (WHT).

\begin{figure}
\includegraphics[angle=-90,width=8cm]{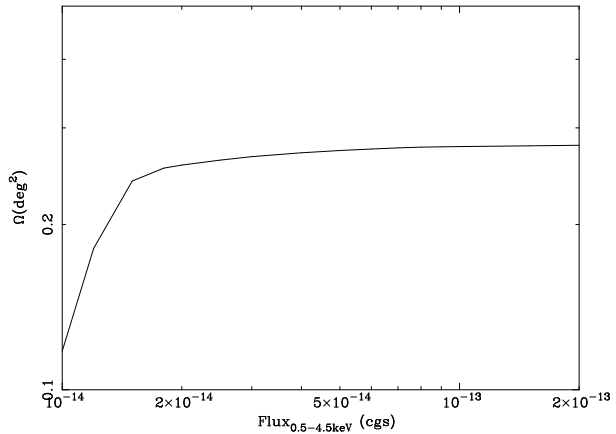}
\caption{Solid angle surveyed as a function of source flux.}
\label{omegasurv}
\end{figure}

In this paper we report on the first results obtained in the AXIS
medium sensitivity survey at high galactic latitude. X-ray sources
serendipitously found in two XMM-Newton observations (field names
\object{G133-69} Pos\_2 and \object{Mkn 205}) have been studied and
followed up down to a 0.5-4.5 keV flux of $2\times 10^{-14}\, {\rm
erg}\, {\rm cm}^{-2}\, {\rm s}^{-1}$. The survey reaches a source
density in excess of 100 sources $\deg^{-2}$, which is appropriate for
spectrocopic follow up using fibre spectroscopy. Our survey is
therefore shallower than the XMM-Newton Lockman Hole survey
\citep{Hasinger2001} and the $Chandra$ deep surveys
\citep{Mushotzky2000,Barger2001,Giacconi2001}.  The surface density
reached is, however, similar to that of the so-called {\it Rosat Deep
Survey} \citep{Boyle1994} and deeper than the $RIXOS$ survey
\citep{Mason2000}.

Although we are still dealing with a small number of sources (29) the
sample presented here provides a flavour of the dominant X-ray source
populations at high galactic latitude down to that flux level. The
paper also describes the observational techniques that we are
following in the AXIS project to build up larger source catalogues at
various flux levels and galactic latitudes.

\section{The X-ray data}

\subsection{XMM-Newton observations}

The identifications presented in this paper correspond to X-ray
sources serendipitously found in 2 XMM-Newton images (\object{G133-69} Pos\_2
and \object{Mkn 205}). \object{G133-69} Pos\_2 was observed as a Guaranteed Time
observation to probe the galactic halo using X-ray shadows. It
consists of a single data set totalling 16 ks of good exposure time in
full frame mode.  The \object{Mkn 205} field was observed as a calibration
observation, and consists of 3 exposures of 17 ks each. One of these
is in large-window mode (for the EPIC pn camera) which only covers
half of the field of view.  The remaining 2 data sets are in full
window mode, and we merged them. One of these was free of particle
background flares, but the other one was strongly contaminated and
only $< 3$ ks of it survived the cleaning process. Details of the
X-ray observations are reported in Table \ref{XMMobs}.

\begin{table}
\caption{Details of XMM-Newton observations.}
\begin{tabular}{ l l l}
Target & \object{G133-69} Pos\_2 & \object{Mkn 205}\\ \hline
Observation date & 03-07-2000 & 06-05-2000         \\
XMM-Newton Obsid & 0112650501 & 0124110101 \\
RA(J2000) & 01:04:00 & 12:21:44  \\
DEC(J2000) & -06:42:00 & 75:18:37 \\
$ b (\deg)$ &  -69.35 &  +41.67  \\
Clean exposure time (ks)$^a$ & 15.86 & 18.97  \\
pn Filter &  Thin & Medium  \\
$N_{HI}$ & $5.17\times 10^{20}$ & $2.81\times 10^{20}$ \\
Conversion factor$^b$ & $2.47\times 10^{-12}$ &  $2.39\times 10^{-12}$ \\
\hline
\end{tabular}

$^a$ This corresponds to the maximum of the exposure map in the pn image

$^b$ The conversion factor is
the ratio between flux in ${\rm erg}\, {\rm cm}^{-2}\, {\rm s}^{-1}$
and pn count rate, both in the 0.5-4.5 keV band.
\label{XMMobs}
\end{table}

Both of these observations have been processed through the pipeline
processing system\footnote{See
http://xmmssc-www.star.le.ac.uk} \citep{Watson2001}, using tasks from
the XMM-Newton Science Analysis Software (SAS) v5.1. Subsequent
analyses were also performed using the same version of the SAS. All
event patterns (single, double and triple) were kept when constructing
the event files.  This provides maximum sensitivity at high photon
energies but the fraction of non-X-ray events rejected is consequently
smaller. When flaring intervals are removed from the event lists,
higher S/N is usually achieved by keeping all patterns.

The pipeline processing source searching procedure has been adopted. We
now sketch briefly its main features. We have used data from EPIC-pn because its
sensitivity is double that of the invidual EPIC-MOS detectors.  Images were
extracted in the following 4 spectral bands: 0.5-2 keV, 2-4.5 keV,
4.5-7.5 keV and 7.5-10 keV. Exposure maps, which account for vignetting,
CCD gaps, bad columns and bad pixels, were constructed for each band.
The combination of the 4 images was used to search for sources with an
overall likelihood above 16. This corresponds to a probability of a
source being spurious of $\sim 8\times 10^{-5}$, i.e., up to one spurious
detection per image with that likelihood. 

\begin{figure}
\includegraphics[angle=-90,width=8cm]{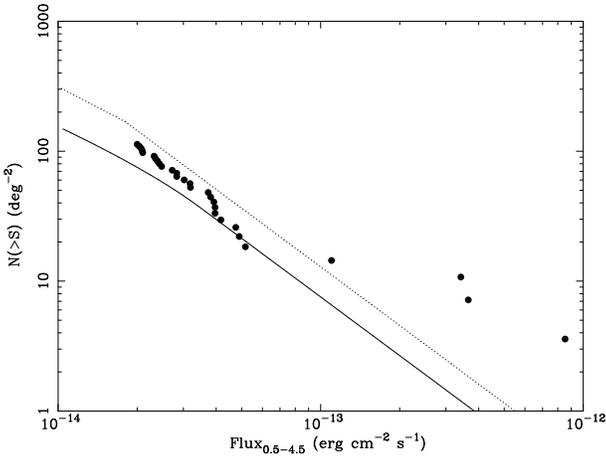}
\caption{Log N-Log S X-ray flux relation for this sample in the
0.5-4.5 keV band (filled circles). For comparison we overlay source
counts from $ROSAT$ (continuous curve) from \citet{Hasinger1998}
converted with a $\Gamma=2$ spectrum, and from $ASCA$ \citep{Ueda1999}
and $Chandra$ \citep{Mushotzky2000} in the 2-10 keV band (merged
together, dotted curve), converted with a $\Gamma=1.6$ power-law
spectrum.}
\label{logNlogS}
\end{figure}

First, a simple sliding box algorithm was applied to mask out the
brightest sources and to spline-fit the background in each CCD chip.
The sliding box algorithm was applied again to search for sources
significant against the fitted background.  Finally a maximum
likelihood fit of the source profiles to the images, simultaneous
to all bands, was performed to produce a final source list with
exposure corrected count rates in each band. Sources were sorted in
terms of the flux in the 0.5-4.5 keV flux. Table \ref{XMMobs} lists
the countrate to flux conversion factors for both fields which we
derived assuming a standard $\Gamma=1.7$ power law absorbed by the
galactic column. Fluxes are corrected for galactic absorption. We
experimented with variations in the spectral shape of the sources and
found changes of up to 15\% only in the conversion factor when varying
the spectral index from $\Gamma=1.5$ to $\Gamma=2$.  That was expected
as the 0.5-4.5 keV band was selected because of the fairly flat
sensitivity of XMM-Newton accross the whole band. As there are still
issues regarding the processing and calibration of EPIC data
(out-of-time events, multiple pattern events, etc.), our listed fluxes
have to be understood modulo these uncertainties.

For each 0.5-4.5 keV image we also extracted a sensitivity map (using
the SAS task {\tt esensmap}) showing the minimum count rate that a
point source must have to be detectable by the algorithm used, at every
position of the detector.  We choose a likelihood limit of 10 (as
opposed to 16) since we are dealing with a single band. This gives a
probability $\sim 4.5\times 10^{-5}$ for a spurious detection in a
single band (again $\sim 1$ spurious detection per image), similar to
the multi-band source search. This map takes into account vignetting,
inter-CCD gaps and bad pixels and columns. We then computed the total
solid angle where a source brighter than any given flux limit would
have been detected. In the case of the \object{Mkn 205} we further
excluded a circle of radius 2 arcmin around the target. The field also
happens to contain an extended source which we identified as the
galaxy \object{NGC 4291}. The extent of this X-ray source effectively
masks out a circle of 1.5 arcmin radius around it where the much
enhanced background due to extended X-ray emission prevents us from
detecting any further X-ray sources. We therefore included this galaxy
as a serendipitous source in our survey but ignored any other possible
sources within a circle of 1.5 arcmin radius around
it. Fig.~\ref{omegasurv} shows the solid angle surveyed at each flux
level. At $2\times 10^{-14}\, {\rm erg}\, {\rm cm}^{-2}\, {\rm
s}^{-1}$, which is our survey limit, we have surveyed 0.26~$\deg^2$.

\subsection{X-ray sources}

The source-searching algorithms produced a number of sources in each
field (37 and 52 in the \object{G133-69} Pos\_2 and \object{Mkn 205} data respectively), some as
faint as $5\times 10^{-15}\, {\rm erg}\, {\rm cm}^{-2}\, {\rm s}^{-1}$
(all fluxes refer to the 0.5-4.5 keV band). The faintest of these
sources have likelihood detections very close to our threshold.  When
truncating at a flux of $2\times 10^{-14}\, {\rm erg}\, {\rm cm}^{-2}\,
{\rm s}^{-1}$ a total of 12 and 17 X-ray sources were found in the
\object{G133-69} Pos\_2 and \object{Mkn 205} fields respectively. Two sources in the \object{Mkn 205} field
whose positions fall in CCD gaps have already been excluded from the
list.  The sensivity of both fields was clearly enough to ensure
completeness at that flux level, i.e., sources significantly fainter
than our survey limit were detected in each field.  Care was taken to
visually screen the source lists in order to clean it from artifacts
derived from the proximity of inter-CCD gaps, bad columns or other
cosmetic effects. 

The X-ray source list was then astrometrically corrected. The
pipeline-processed data provides astrometry drawn from the attitude
and orbit control system (AOCS), which we believe to be good to within
a few arcsec.  To further improve that, we cross-correlated the source
positions of the X-ray sources obtained with a list of optical sources
obtained from the i'-band images (see Sect. \ref{opticalimaging})
using the SAS task {\tt eposcorr}.  To this goal, we used all detected
X-ray sources and all detected optical sources in the field to
maximize the number of matches. The situation is summarized in Table
\ref{tab-eposcorr}, were we show that significant astrometric shifts
were still present in both data sets. One point that needs stressing
is that the number of X-ray to optical matches is much larger when we
use the full catalogue of optical sources obtained in our wide-field
images than if we used, e.g., the USNO A2 \citep{Monet1998} catalogued
sources.  In that case we would be restricted to $\sim 10$ matches per
field, with the corresponding uncertainty in the astrometric
correction parameters.  We further discuss the accuracy of the
astrometric solution in the X-ray images in Sect.
\ref{astrometry}. Table \ref{XMMsources} lists the X-ray sources
brighter than our survey limit with the astrometrically corrected
positions.

\begin{table}
\caption{Astrometric correction to pipeline processed XMM-Newton data,
derived from cross-correlation with the i'-band optical images}
\begin{tabular}{l c c c c}
\hline
Field    & $\Delta RA (\arcsec)$ & $\Delta DEC (\arcsec)$
& Rotation ($\deg$) \\ \hline
\object{Mkn 205}   & $-2.67$ & $+3.09$ &
$-0.16$ \\
\object{G133-69} Pos\_2 & $-0.73$  & $+0.63$  &
$-0.35$  \\
\hline
\end{tabular}
\label{tab-eposcorr}
\end{table}

In order to gain X-ray spectral information on the X-ray sources, we
used the count rates in individual bands to construct the following
hardness ratios:

\[
HR_1={C(2-4.5\, {\rm kev})-C(0.5-2\, {\rm kev})\over C(2-4.5\, {\rm
kev})+C(0.5-2\, {\rm kev})}
\]
and
\[
HR_2={C(4.5-10\, {\rm kev})-C(2-4.5\, {\rm kev})\over C(4.5-10\, {\rm
kev})+C(2-4.5\, {\rm kev})}
\]

Note that these are computed using the exposure-map corrected
count-rates, so that energy dependent vignetting is approximately
corrected.  All sources in the sample selected here had positive count
rates in all bands.Also, note that the typical number of counts per source is
fairly small and therefore a detailed individual spectral analysis of
each source is difficult and beyond the scope of this paper.  Table
\ref{XMMsources} contains the basic X-ray data on these sources.

\begin{sidewaystable*}
\caption{X-ray sources serendipitously discovered in the XMM-Newton
fields under study} 
\begin{tabular}{l c c c c c c l}
Source name & RA$_{\rm X}^a$ (J2000) & DEC$_{\rm X}^a$ (J2000) &Perr$^b$ & Flux$^c$ & $HR_1$ & $HR_2$ & Comments\\ \hline
\object{XMMU J010316.7-065137}&  01:03:16.72&-06:51:37.28&1.99 &  $2.32\pm 0.43$ & $-0.62\pm  0.17$ &  $-0.01\pm  1.00$&\\ 
\object{XMMU J010327.3-064643}&  01:03:27.30&-06:46:43.69&0.80 &  $4.15\pm 0.37$ & $-0.15\pm  0.01$ &  $-0.29\pm  0.37$&Close to pn noisy column\\ 
\object{XMMU J010328.7-064633}&  01:03:28.71&-06:46:33.34&1.30 &  $2.72\pm 0.31$ & $-0.79\pm  0.08$ &  $+0.09\pm  1.00$&Close to pn noisy column\\ 
\object{XMMU J010333.8-064016}&  01:03:33.86&-06:40:16.07&0.82 &  $2.65\pm 0.24$ & $-0.53\pm  0.08$ &  $-0.03\pm  0.17$&\\ 
\object{XMMU J010339.8-065224}&  01:03:39.87&-06:52:24.74&0.96 &  $3.73\pm 0.39$ & $-0.74\pm  0.08$ &  $+0.23\pm  0.27$&\\ 
\object{XMMU J010355.6-063710}&  01:03:55.62&-06:37:10.48&0.82 &  $2.56\pm 0.22$ & $-0.47\pm  0.08$ &  $-0.91\pm  1.00$&\\ 
\object{XMMU J010400.9-064949}&  01:04:00.96&-06:49:49.37&0.96 &  $2.70\pm 0.27$ & $-0.91\pm  0.09$ &  $+0.25\pm  1.00$&\\ 
\object{XMMU J010410.5-063926}&  01:04:10.56&-06:39:26.46&0.62 &  $4.17\pm 0.28$ & $-0.65\pm  0.05$ &  $-0.64\pm  0.65$&\\ 
\object{XMMU J010420.9-064701}&  01:04:20.91&-06:47:01.46&0.89 &  $2.39\pm 0.25$ & $-0.62\pm  0.09$ &  $-0.13\pm  0.27$&\\ 
\object{XMMU J010430.1-064456}&  01:04:30.13&-06:44:56.07&0.72 &  $4.90\pm 0.36$ & $-0.53\pm  0.06$ &  $-0.46\pm  0.53$&\\ 
\object{XMMU J010437.5-064739}&  01:04:37.56&-06:47:39.29&1.15 &  $2.29\pm 0.30$ & $-0.65\pm  0.12$ &  $+0.27\pm  0.27$&\\ 
\object{XMMU J010444.6-064833}&  01:04:44.68&-06:48:33.42&1.16 &  $3.96\pm 0.43$ & $-0.63\pm  0.09$ &  $-0.16\pm  0.34$&\\ 
\object{XMMU J121819.4+751919}&  12:18:19.48&+75:19:19.61&1.29 &  $3.17\pm 0.41$ & $-0.79\pm  0.02$ &  $-0.38\pm  1.00$&\\ 
\object{XMMU J122017.9+752212}&  12:20:17.98&+75:22:12.17&0.28 &  $89.6\pm 0.72$ & $-0.90\pm  0.01$ &  $-0.66\pm  0.11$&Extended$^d$\\ 
\object{XMMU J122048.4+751804}&  12:20:48.43&+75:18:04.10&0.69 &  $2.97\pm 0.24$ & $-0.73\pm  0.02$ &  $-0.35\pm  0.32$&\\ 
\object{XMMU J122051.4+752821}&  12:20:51.45&+75:28:21.84&1.26 &  $2.07\pm 0.30$ & $-0.76\pm  0.03$ &  $+0.26\pm  0.51$&\\ 
\object{XMMU J122052.0+750529}&  12:20:52.02&+75:05:29.44&0.40 &  $36.0\pm 1.31$ & $-0.60\pm  0.01$ &  $-0.33\pm  0.05$&\\ 
\object{XMMU J122111.2+751117}&  12:21:11.29&+75:11:17.19&0.54 &  $4.12\pm 0.31$ & $-0.66\pm  0.01$ &  $-0.40\pm  0.66$&\\ 
\object{XMMU J122120.5+751616}&  12:21:20.56&+75:16:16.10&0.57 &  $2.98\pm 0.23$ & $-0.84\pm  0.01$ &  $-0.50\pm  0.41$&\\ 
\object{XMMU J122135.5+750914}&  12:21:35.59&+75:09:14.28&0.63 &  $4.99\pm 0.36$ & $-0.69\pm  0.02$ &  $-0.14\pm  0.51$&\\ 
\object{XMMU J122143.8+752235}&  12:21:43.88&+75:22:35.32&0.77 &  $2.20\pm 0.21$ & $-0.57\pm  0.04$ &  $-0.56\pm  0.11$&\\ 
\object{XMMU J122206.4+752613}&  12:22:06.48&+75:26:13.78&0.21 &  $38.4\pm 0.96$ & $-0.52\pm  0.01$ &  $-0.35\pm  0.02$&\\ 
\object{XMMU J122242.6+751434}&  12:22:42.69&+75:14:34.96&0.75 &  $2.15\pm 0.21$ & $-0.45\pm  0.05$ &  $-0.56\pm  0.21$&Close to pn CCD gap\\ 
\object{XMMU J122258.1+751934}&  12:22:58.11&+75:19:34.31&0.55 &  $5.44\pm 0.36$ & $-0.44\pm  0.03$ &  $-0.49\pm  0.06$&Close to pn CCD gap\\ 
\object{XMMU J122318.5+751504}&  12:23:18.58&+75:15:04.08&0.55 &  $3.35\pm 0.31$ & $-0.72\pm  0.02$ &  $-0.43\pm  0.42$&\\ 
\object{XMMU J122344.7+751922}&  12:23:44.79&+75:19:22.18&0.60 &  $3.34\pm 0.24$ & $-0.74\pm  0.02$ &  $-0.82\pm  0.28$&\\ 
\object{XMMU J122351.0+752227}&  12:23:51.02&+75:22:27.99&0.36 &  $11.6\pm 0.56$ & $-0.62\pm  0.01$ &  $-0.44\pm  0.05$&\\ 
\object{XMMU J122435.7+750812}&  12:24:35.77&+75:08:12.02&1.36 &  $2.10\pm 0.37$ & $-0.77\pm  0.04$ &  $-0.20\pm  1.00$&\\ 
\object{XMMU J122445.5+752224}&  12:24:45.53&+75:22:24.80&0.88 &  $4.16\pm 0.42$ & $-0.77\pm  0.02$ &  $-0.31\pm  0.73$&\\ 
\hline
\end{tabular}

Notes to Table: $^a$ Position of the X-ray source after astrometric
correction; $^b$ Position error in arcsec (statistical only); $^c$
Flux in the 0.5-4.5 keV band in units of $10^{-14}\, {\rm erg}\, {\rm
cm}^{-2}\, {\rm s}^{-1}$; $^d$ The flux for this source has been
estimated from aperture photometry.
\label{XMMsources}
\end{sidewaystable*}

In Fig. \ref{logNlogS} we plot the integrated source counts derived
from this survey. The source density at the survey limit is $113\pm
21$ sources $\deg^{-2}$. This curve is to be compared with the $ROSAT$
source counts \citep{Hasinger1998} which we converted to our
0.5-4.5 keV band by using a $\Gamma=2$ spectrum which is consistent
with the average value of $HR_1$ (see below).  Our source counts are
$\sim 60\%$ higher than the $ROSAT$ ones at the survey limit.
Choosing a conversion factor corresponding to a power-law with
spectral index $\Gamma=1.7$ just brings the $ROSAT$ curve up by $\sim
15\%$, still significantly below our source counts.  

We also compare the source counts with those derived in the 2-10 keV
from ASCA \citep{Ueda1999} and $Chandra$ \citep{Mushotzky2000}. Note
that the \citet{Baldi2001} source counts, based on XMM-Newton data,
are consistent with these earlier work. Here a conversion factor for
$\Gamma=1.6$ has been applied, which is consistent with the value of
$HR_2$ (see below). Our source counts fall modestly below (17\%, only
$\sim 1$ sigma significant) those in the harder band at the survey
limit.  It is clear that by selecting in the 0.5-4.5 keV band we find
a large fraction of the sources that $ROSAT$ missed, but we may still
be missing a small fraction of the hard sources detected in the 2-10
keV band.

\begin{figure}
\resizebox{\hsize}{!}{\includegraphics{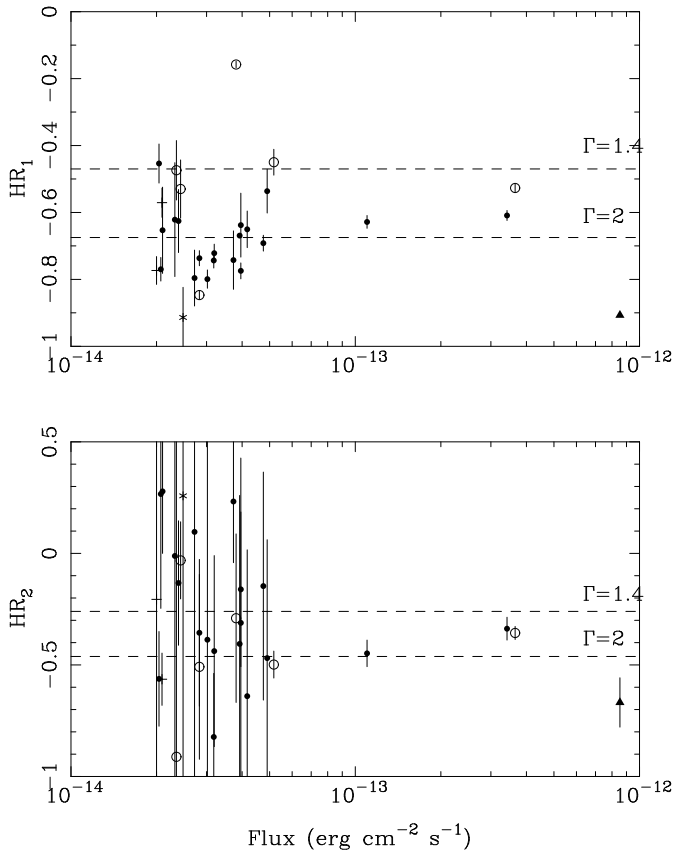}}
\caption{Hardness ratios $HR_1$ and $HR_2$ as a function of X-ray
flux. Symbols are as follows: filled dots are BLAGNs, empty circles
are NELGs, triangles are Galaxies, asterisks are AC and crosses are
non-identified sources. We also overlay expected values of $HR_1$ and
$HR_2$ for single power-law spectra with $\Gamma=2$ and $\Gamma=1.4$}
\label{HRs}
\end{figure}

\begin{figure}
\resizebox{\hsize}{!}{\includegraphics{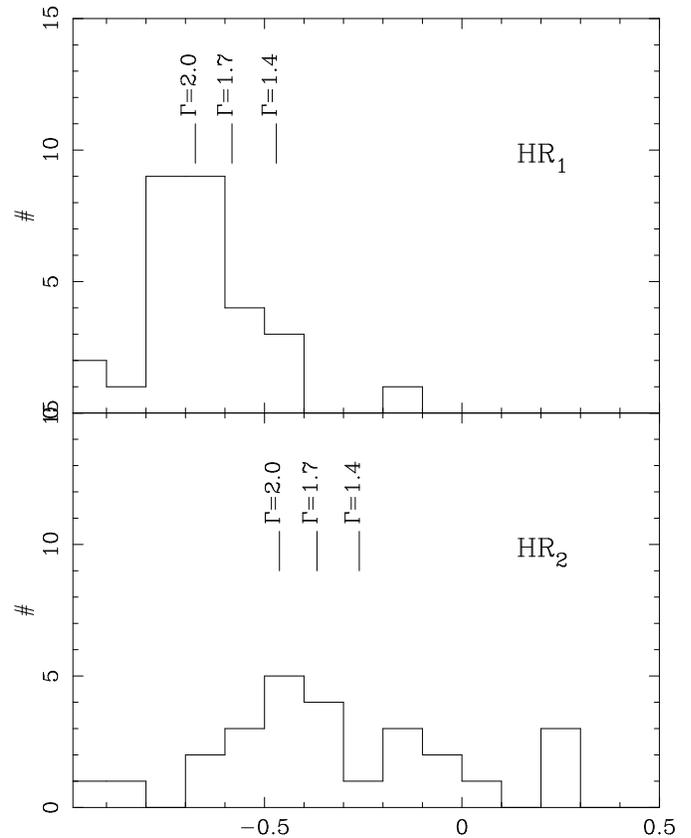}}
\caption{Histograms of the values of $HR_1$ and $HR_2$. Marks for standard power-law spectra with various values of the 
photon spectral index $\Gamma$ are shown.}
\label{histo-HRs}
\end{figure}

Fig. \ref{HRs} shows the hardness ratios for the sources as a function
of flux. We must caution that although the two fields under
consideration have different Galactic absorbing columns and
furthermore have been observed with different EPIC-pn filters (thin
and medium, see Table \ref{XMMobs}), the
effect of both of these facts on the hardness ratios (particularly
$HR_1$) is completely negligible. No obvious trend of the source
spectra is seen with source flux.  The so-called spectral paradox
\citep{Fabian1992}, i.e., the source's spectra being steeper than the
X-ray background spectrum ($\Gamma\sim 1.4$), is therefore not
obviously solved at this flux level, i.e., we do not find a
significant sample of sources that have an X-ray spectrum close to
that of the X-ray background. This is also revealed by the histograms
of $HR_1$ and $HR_2$ which are shown in fig. \ref{histo-HRs} together
with the expected values for various power laws with galactic
absorption. The weighted averages of the hardness ratios are $\langle
HR_1\rangle=-0.69\pm 0.03$ and $\langle HR_2\rangle=-0.37 \pm 0.03$,
consistent with a fairly steep X-ray spectrum
($\Gamma=1.6-2.0$). Further details on the X-ray spectra of the
different source classes are given in Sect. \ref{global}.

\section{Optical observations}

\subsection{Imaging}\label{opticalimaging}

Both target fields were observed with the Wide-Field Camera (WFC) on
the 2.5m INT telescope in dark time. Table \ref{WFCobs} summarizes
the main features of the images obtained. The WFC covers virtually all
the field of view of EPIC, if centered optimally. In our case, all
sources listed in Table \ref{XMMsources} were imaged.

\begin{table}
\caption{Details of the optical wide-field imaging performed with the
WFC on the INT.}
\begin{tabular}{ l l l}
Target & \object{Mkn 205} & \object{G133-69} Pos\_2 \\ \hline
Observation date & 30-04-2000 & 25/26-07-2000\\
Seeing (arcsec in i') & 1.5 & 1.1  \\
i' limiting magnitude & 22.1   & 23.0  \\
u exposure time (s) & 600 & -  \\ 
g' exposure time (s) & 600 & 600  \\ 
r' exposure time (s) & 600 & 600  \\ 
i' exposure time (s) & 1200 & 1200  \\ 
Z exposure time (s) & 1200 & 1200  \\ \hline
\end{tabular}
\label{WFCobs}
\end{table}

The WFC images were reduced using standard techniques including
de-bias, non-linearity correction, flat fielding and fringe correction
(in i' and Z). Bias frames and twilight flats obtained during the same
observing nights were used, but for the fringe correction
contemporaneous archival i' and Z fringe frames were utilised. Information
on the WFC pipeline procedures, which performs all these steps can be
found under the Cambridge Astronomy Survey
Unit\footnote{http://www.ast.cam.ac.uk/\~{}wfcsur} (CASU) web pages.

The photometric calibration was performed by assuming average extinction
constants and archival zero-point constants obtained routinely with
the WFC, rather than measuring both from photometric standards.

Astrometric calibration was performed in two different ways. First a
12 parameter plate solution was applied to each one of the 4 CCDs of
the WFC independently, by matching sources found in the image to USNO
source positions. Residuals were typically found to be of the order of
0.3 arcsec for $\sim 100-200$ matched sources.  A second astrometric
calibration (along the lines of the WFC survey) was performed by a
simple 6 parameter plate solution over the whole 4-CCD image, but
taking into account the previously calibrated telescope distortion
cubic term.  In this case the sources found were matched to APM source
positions with similar residuals.  We found the position of the
candidate counterparts to X-ray sources to be consistent in both
methods to within $<0.5$~arcsec.

\subsection{Selection of candidate counterparts}

In order to search for candidate counterparts of the X-ray sources, we
have used the i'-band WFC image.  Optical source lists for
these images have been constructed by using the Sextractor algorithm
\citep{Bertin1996}. Sources were recorded if a minimum number of
10 connected pixels (of 0.33 arcsec) lay above 2 standard deviations
of the background. Indeed some very faint sources escaped this
detection algorithm, but the impact on the search for candidate
counterparts was very limited.

Counterparts for the X-ray sources were searched for in the optical
image lists. Candidate counterparts had to be either within the
5$\sigma$ statistical error or within 5 arcsec from the position of
the X-ray source. This last criterion was used to accommodate any
residual systematics in the astrometric calibration of the X-ray EPIC
images. 

The result of this search was very encouraging. Out of the 29 X-ray
sources, 24 showed a single candidate counterpart, 4 showed two
candidate counterparts and the remaining source showed no or very
faint candidate counterpart in the i'-band image. The two basic
conclusions to extract from this are that the selection of filter and
depth of the images are appropriate to this sample and that the X-ray
positional errors provided by XMM-Newton are good enough to single out
a unique counterpart for the vast majority of the high-galactic
latitude sources.

\subsection{Spectroscopic observations}

Candidate counterparts were spectroscopically observed with various
spectrographs in order to identify the nature of the sources. Table
\ref{SPECobs} lists the relevant parameters of the spectroscopic
instrumental setup used.  At the source density we are dealing with
($\sim 100\deg^{-2}$) fibre spectrographs are likely to be the most
efficient. However, we had a very limited amount of nights with this
instrumentation and furthermore the fibres used were too large (2.7
arcsec diameter) to detect the faintest objects. Therefore a
significant fraction of the sources were identified with long-slit
spectrographs.

\begin{table*}
\caption{List of spectroscopic setups relevant to this sample.}
\begin{tabular}{r c c c c l}
\hline
Telescope & Instrument       & Spectral       & Slit width  & Spectral              & Comments \\
          &                  & range (\AA )   & (arcsec)    & resolution$^a$ (\AA ) & \\ \hline 
 WHT      & AUTOFIB2/WYFFOS  & 3900-7100      &  2.7$^b$    & 6-7                   & Fibre  \\
 WHT      & ISIS             & 3500-8500      &  1.2-2.0    & 3.0-3.3               & Long slit  \\
 TNG      & DOLORES          & 3500-8000      &  1.0-1.5    & 14-15                 & Long slit  \\
\hline
\end{tabular}

$^a$ Measured from unsaturated arc lines\\ 
$^b$ Width of individual fibres 
\label{SPECobs}
\end{table*}

As pointed out before, the vast majority of X-ray sources have a
unique candidate counterpart. For the cases where there were two
candidate counterparts we observed the brightest one (which happened
also to be the closest one to the X-ray source) which
invariably turned out to be a plausible identification (i.e. some sort
of AGN). 

The spectra were reduced using standard IRAF techniques. These
included de-biasing, flat fielding, illumination correction (whenever
a twilight flat was available), cosmic ray rejection, spectral
extraction and background subtraction, arc lamp wavelength calibration
and flux calibration by using spectrophotometric standards. The flux
calibration does not have to be understood in absolute terms, as
significant fractions of the light escaped the corresponding apertures
and no attempt has been made to correct for this. Nevertheless the
overall spectral shape, if not the normalisation, should be
approximately correct. That is especially true for the long-slit
spectra which in the majority of the cases were obtained with the slit
oriented in parallactic angle. 

The case of fibre spectroscopy deserves further comment. The
relative fibre throughput was obtained by observing in offset sky
positions within the same fibre configuration under use. In the proper
on-target observations, the sky was subtracted by combining all the
sky fibres that we placed in regions free from bright sources. We
applied the same procedure to the offset sky observations, which
allowed us to tweak the relative sensitivity of the fibres by making
sure that no significant residuals were left in these spectra after
the sky was subtracted. In a couple of iterations we found a
satisfactory solution for the relative fibre sensitivity and then we
used this to sky-subtract the target apertures. It must be stressed
that sky lines are very difficult to subtract at this spectral
resolution and therefore residuals will unavoidably be present in the
fibre spectra.

We must emphasize that fibre spectroscopy, even with the large fibre
aperture of 2.7 arcsec, was very efficient in identifying the
optically brightest conterparts in the \object{Mkn 205} field.  We allocated a
total of 26 fibres on optical sources in a single setup, which
included several corresponding to X-ray sources below the flux
cutoff. We succeeded in identifying 8 X-ray sources above the flux
cutoff and a futher 3 below it in a total time of 2.5 hours including
all overheads. This is at least twice as efficient
as long-slit spectroscopy on the same telescope.

This high efficiency in the fibre spectroscopy relied, however, on two
basic facts: the night was dark (so the sky background and noise were
as low as possible) and the number of X-ray sources with a single
candidate counterpart was very large. It must be pointed out that fibre
spectrographs do not usually allow to place fibres within several
arcsec at best, and therefore it is not possible to observe different
candidate counterparts of the same X-ray source within the same fibre
setup.

\section{Identification of the X-ray sources}

Table \ref{IDs} lists the optical identification of the sources
together with their optical magnitudes. The classification we have
chosen includes several source classes: Broad-Line Active Galactic
Nuclei (BLAGN), Narrow Emission Line Galaxies (NELG, which invariably
turned out to be Narrow-Line AGN), non-emission line galaxies (Gal),
Active Coronal stars (AC), stars without signs of activity (STAR) and
clusters of galaxies (Clus).

\begin{sidewaystable*}
\caption{Optical identifications  of the X-ray sources. Under the last
column we quote the main spectral features detected in the spectra.}
\begin{tabular}{l c c c c c c c l l r l}
Source name & RA$_{\rm O}^a$ (J2000) & DEC$_{\rm O}^a$ (J2000) & u & g' & r' & i' & Z & Class & $z$ & $L_{44}^b$ &  Comments\\ \hline
\object{XMMU J010316.7-065137}&  01:03:16.43&-06:51:35.83&      & 19.85& 19.23 &18.74& 18.57& BLAGN &    1.914&  4.221 & CIII], CIV, MgII\\
\object{XMMU J010327.3-064643}&  01:03:27.41&-06:46:43.31&      & 22.26& 21.86 &21.77& 21.06& NELG? &    1.010?&  2.036 & CIII]?,[OII]?, [NeV]?\\
\object{XMMU J010328.7-064633}&  01:03:28.67&-06:46:32.02&      & 20.26& 19.81 &19.27& 19.15& BLAGN &    1.820&  4.467 & SiIV, CIV, CIII], MgII\\
\object{XMMU J010333.8-064016}&  01:03:33.88&-06:40:16.21&      & 22.70& 21.12 &19.76& 19.22& NELG  &    0.692&  0.597 & [OII],H$\epsilon$?,H$\gamma$,[OIII]\\
\object{XMMU J010339.8-065224}&  01:03:39.89&-06:52:25.99&      & 21.39& 20.65 &20.23& 19.99& BLAGN &    1.128&  2.297 & [NIII],CIII], MgII,[OII]\\
\object{XMMU J010355.6-063710}&  01:03:55.67&-06:37:10.88&      & 21.82& 21.00 &20.48& 20.11& NELG? &    0.314&  0.114 & H$\beta$, [OIII], H$\delta$, [NeV]?\\
\object{XMMU J010400.9-064949}&  01:04:01.09&-06:49:50.89&      & 17.08& 15.52 &15.24& 14.16& AC/dMe&         &        & Ca H\&K, H$\beta$, H$\alpha$\\
\object{XMMU J010410.5-063926}&  01:04:10.54&-06:39:26.69&      & 19.88& 19.47 &19.08& 19.03& BLAGN &    0.630&  0.774 & MgII, H$\gamma$, H$\beta$, [OIII]\\
\object{XMMU J010420.9-064701}&  01:04:20.97&-06:47:01.97&      & 19.71& 18.98 &18.41& 18.34& BLAGN &    1.536&  2.773 & CIV, CIII], MgII\\
\object{XMMU J010430.1-064456}&  01:04:30.07&-06:44:56.60&      & 18.77& 18.27 &18.06& 17.96& BLAGN &    0.910&  1.937 & CIII], MgII, [OII], H$\gamma$, [NeV]\\
\object{XMMU J010437.5-064739}&  01:04:37.55&-06:47:37.23&      & 21.19& 20.71 &20.24& 19.76& BLAGN &    2.511&  7.248 & Ly$\alpha$, SiIV, CIV?, CIII]\\
\object{XMMU J010444.6-064833}&  01:04:44.71&-06:48:33.31&      & 20.52& 20.32 &19.96& 19.58& BLAGN &    2.256&  10.080&  Ly$\alpha$, CIV, CIII]\\
\object{XMMU J121819.4+751919}&  12:18:19.06&+75:19:22.02&19.53 & 19.79& 19.21 &18.96& 18.55& BLAGN &    2.649&  11.209& Ly$\alpha$, CIV, CIII]\\
\object{XMMU J122017.9+752212}&  12:20:17.70&+75:22:18.00&      & 12.27$^c$& 11.70$^c$ & &  & Gal   &   0.0058&  0.0013& \object{NGC 4291}$^d$\\
\object{XMMU J122048.4+751804}&  12:20:48.25&+75:18:07.29&18.41 & 18.97& 18.47 &17.96& 17.83& BLAGN &    1.687&  4.181 & CIV, CIII], MgII\\
\object{XMMU J122051.7+752821}&  12:20:51.73&+75:28:20.67&20.90 & 22.05& 21.33 &20.80&      & BLAGN &    0.181&  0.030 & H$\beta$, H$\gamma$, H$\delta$? OII?\\
\object{XMMU J122052.0+750529}&  12:20:51.27&+75:05:31.62&18.24 & 18.62& 18.20 &17.92& 17.85& BLAGN &    0.646&  7.036 & [OII],MgII\\
\object{XMMU J122111.2+751117}&  12:21:10.62&+75:11:19.39&18.78 & 19.54& 18.79 &18.47& 18.32& BLAGN &    1.259&  3.179 & EMSS source$^e$\\
\object{XMMU J122120.5+751616}&  12:21:19.90&+75:16:18.09&20.21 & 20.90& 19.98 &19.45& 18.85& NELG  &    0.340&  0.156 & [OII], [OIII], [NeV]\\
\object{XMMU J122135.5+750914}&  12:21:34.92&+75:09:15.99&19.84 & 20.38& 19.11 &18.32& 17.77& BLAGN &    0.330&  0.246 & H$\beta$, [OIII]\\
\object{XMMU J122143.8+752235}&             &            &      &      &       &     &      &       &         &        &                       \\
\object{XMMU J122206.4+752613}&  12:22:06.66&+75:26:15.36&20.23 & 20.03& 18.68 &17.83& 17.41& NELG  &    0.238&  0.975 & [OII],[OIII]\\
\object{XMMU J122242.6+751434}&  12:22:42.69&+75:14:34.68&20.97 & 21.82& 21.13 &20.71& 19.98& BLAGN &    1.065&  1.174 & MgII, CIII]?\\
\object{XMMU J122258.1+751934}&  12:22:58.00&+75:19:34.66&      & 22.66& 21.66 &21.10&      & NELG? &    0.257&  0.161 & [OII],[OIII]\\
\object{XMMU J122318.5+751504}&  12:23:18.11&+75:15:04.64&20.27 & 21.15& 20.62 &20.24&      & BLAGN &    1.509&  3.752 & CIII], MgII\\
\object{XMMU J122344.7+751922}&  12:23:45.65&+75:19:23.01&20.11 & 20.69& 20.26 &19.88& 19.53& BLAGN &    0.757&  0.904 & MgII, [NeV]?, [OII]\\
\object{XMMU J122351.0+752227}&  12:23:50.87&+75:22:28.57&19.51 & 19.93& 19.48 &18.76& 18.50& BLAGN &    0.565&  1.718 & MgII, [NeV], [OII], H$\gamma$, [OIII]\\
\object{XMMU J122435.7+750812}&  12:24:35.31&+75:08:10.10&      &      &       &     &      &       &         &        &            \\
\object{XMMU J122445.5+752224}&  12:24:45.42&+75:22:24.92&19.27 & 20.16& 19.82 &19.33& 19.36& BLAGN &    1.852&  7.088 & CIV, CIII]\\
\hline
\end{tabular}

Notes to Table: $^a$ Position of the optical source;
$^b$ Luminosity in the 0.5-4.5 keV rest frame in units of
$10^{44}\, {\rm erg}\, {\rm s}^{-1}$;
$^c$ B and R magnitudes from \citet{deV1991} and \citet{Sandage1978}
quoted under g' and r' respectively;
$^d$ Identification from \citet{Humasson1956};
$^e$ Identification from \citet{Stocke1991}
\label{IDs}
\end{sidewaystable*}

Finding charts and spectra of these sources can be found under the
AXIS programme web pages (http://www.ifca.unican.es/\~xray/AXIS). Here
we add some notes on individual objects

\begin{figure}
\resizebox{\hsize}{!}{\includegraphics{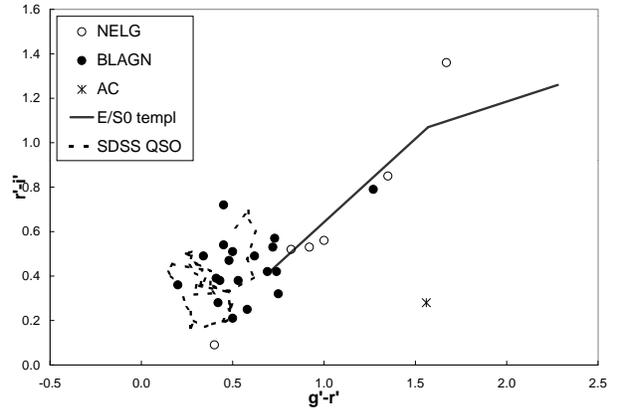}}
\caption{Optical colour-colour diagram. Symbols as in \ref{HRs}. The
continuous line represents the colours of E/S0 galaxies in the
redshift range $z=0-1$ and the dashed line the median colours of SDSS
QSOs in the redshift range $z=0-3$.}
\label{Optcol}
\end{figure}

\object{XMMU J010327.3-064643} is the faintest optical source (in r',
i' and Z) with hints of narrow emission lines. We have binned the
spectrum into bins of 600 \AA\ each, and applied a photometric
redshift technique using a set of galaxy templates as in
\citet{fsoto1999}.  The minimum $\chi^2$ happens at $z=1.01$, which
allows us to interpret an apparently narrow emission line as CIII],
and a dubious [OII].  We therefore tentatively classify this source as
a NELG, although its high luminosity ($1.8 \times 10^{44}\, {\rm
erg}\, {\rm s}^{-1}$) would promote it to QSO2. The optical colours of
this source do indeed correspond to a QSO or to a starforming galaxy
(see fig. \ref{Optcol}, where this source is the empty circle at the
bottom of the graph). The classification and redshift of this source
are at present very uncertain.

\object{XMMU J122258.1+751934} is a faint source, with a detected but weak emission line
that we believe to be [OII] and dubious [OIII] doublet at
z=0.257. Since the spectrum is very noisy, we applied the photometric
redshift technique to the binned spectrum. The minimum
$\chi^2$ is indeed found at $z\sim 0.3$.

\object{XMMU J122120.5+751616} appears to have a very blue spectrum, but the H$\beta$
line, which is weak, does not appear to be broad. On the other hand it
has strong [NeV], [OII] and [OIII] doublet. Therefore we have
classified it as a NELG.

\object{XMMU J010328.7-064633} and \object{XMMU J010316.7-065137} appear to be BAL QSOs. They will be
discussed in Sect. \ref{BALQSOs}

\begin{figure}
\resizebox{\hsize}{!}{\includegraphics{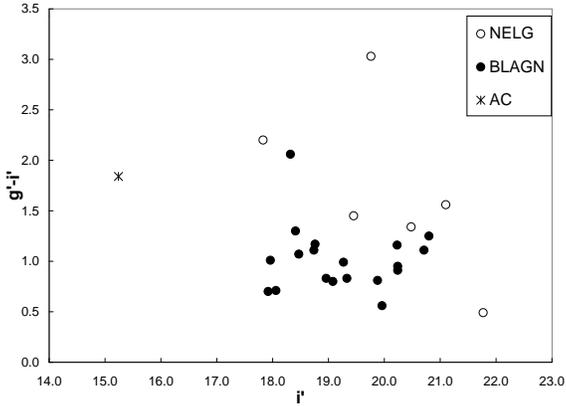}}
\caption{Optical colour $g'-i'$ as a function of i'-magnitude. Symbols as in \ref{HRs}}
\label{gmivsimag}
\end{figure}

Fig. \ref{gmivsimag} shows the $g'-i'$ colour of the identified
sources as a function of their i'-magnitude.  As expected, BLAGNs are
usually bluer than the NELGs where the optical light is dominated
by the host galaxy rather than by the active nucleus.

To further explore the optical colours, we plot in fig. \ref{Optcol}
the $r'-i'$ versus $g'-r'$ optical colour-colour diagram for the
identified sources.  We overlay median colours for QSOs, as derived from the
Sloan Digitized Sky Survey \citep{Richards2001}, in the redshift range
$z=0-3$ and for E/S0 galaxies, as derived from the \citet{Coleman1980}
template, in the redshift range $z=0-1$. For clarity we do not show
the expected colours for starforming galaxies as they are mostly
coincident with those of QSOs, but they can follow the track marked by
the E/S0 colours by adding increasing amounts of reddening. The
influence of cosmological Lyman-$\alpha$ absorption has not been
included, as it is only relevant in the g'-band for redshifts $z>
3$. Clearly, the X-ray sources classified as BLAGNs have optical
colours as expected for QSOs, and the X-ray sources identified as
NELGs fall in the region where the galaxies lie (either E/S0 or more
likely reddened starforming galaxies). This implies again that
in NELGs most of the optical light we see comes from the host galaxy,
unlike the X-ray emission which comes from the nucleus.

\subsection{Additional identifications}

Besides the identifications corresponding to the complete sample of
X-ray selected sources down to an X-ray flux of $2\times 10^{-14}\,
{\rm erg}\, {\rm cm}^{-2}\, {\rm s}^{-1}$, we have identified a few
sources fainter than these.  These 10 additional sources are not part of
the sample considered here, but still we list their identifications to
help other researchers in possible identification work of fainter
sources. These identifications are listed in Table \ref{addIDs}.

\begin{sidewaystable}
\caption{Additional identifications below $2\times 10^{-14}\, {\rm
erg}\, {\rm cm}^{-2}\, {\rm s}^{-1}$}
\begin{tabular}{l c c c c c c c l c r}
Source name & RA$_{\rm X}^a$ (J2000) & DEC$_{\rm X}^a$ (J2000) & Perr$^b$ & Flux$^c$ & RA$_{\rm O}^d$ (J2000) & DEC$_{\rm O}^d$ (J2000) & i' & Class & z & $L_{44}^e$\\ \hline
\object{XMMU J010311.7-064038}&  01:03:11.79&-06:40:38.25& 1.93& $0.83\pm 0.24$& 01:03:11.83& -06:40:39.14& 18.98&  NELG   &0.187 &0.133 \\
\object{XMMU J010359.8-065318}&  01:03:59.82&-06:53:18.46& 1.95& $1.25\pm 0.65$& 01:03:59.87& -06:53:18.87& 20.92&  BLAGN  & 1.270& 0.980\\
\object{XMMU J010400.9-063027}&  01:04:00.98&-06:30:27.36& 2.34& $1.55\pm 0.30$& 01:04:01.15& -06:30:28.35& 20.09&  BLAGN  & 1.693& 2.199\\
\object{XMMU J010402.9-063600}&  01:04:02.91&-06:36:00.14& 1.14& $1.51\pm 0.20$& 01:04:02.84& -06:35:59.70& 21.69&  BLAGN  & 0.932& 0.629\\
\object{XMMU J010405.5-065359}&  01:04:05.57&-06:53:59.21& 2.45& $1.27\pm 0.26$& 01:04:05.33& -06:53:59.17& 21.13&  BLAGN  & 2.821& 5.093\\
\object{XMMU J010410.0-063012}&  01:04:10.06&-06:30:12.92& 1.77& $1.63\pm 0.72$& 01:04:10.13& -06:30:12.78& 21.13&  BLAGN  & 1.190 &1.579\\
\object{XMMU J010411.5-065209}&  01:04:11.53&-06:52:09.60& 2.25& $1.56\pm 0.26$& 01:04:11.40& -06:52:08.86& 20.90&  BLAGN  & 1.224& 1.138\\
\object{XMMU J010439.3-064629}&  01:04:39.34&-06:46:29.88& 1.60& $1.49\pm 0.24$& 01:04:39.33& -06:46:27.24& 20.08&  BLAGN  & 1.620& 1.927\\
\object{XMMU J121937.5+751042}&  12:19:37.57&+75:10:42.96& 1.46& $1.61\pm 0.28$& 12:19:37.34& +75:10:43.79& 20.77&  STAR   &      &      \\
\object{XMMU J122425.4+751818}&  12:24:25.46&+75:18:18.04& 1.34& $1.36\pm 0.22$& 12:24:25.50& +75:18:19.74& 15.72&  AC/dMe &      &      \\
\hline
\end{tabular}

Notes to Table: $^a$ Position of the X-ray source; $^b$ Position error (Statistical)
of the X-ray source in arcsec; $^c$ Flux in the 0.5-4.5 keV band, in units of $10^{-14}\, {\rm
erg}\, {\rm cm}^{-2}\, {\rm s}^{-1}$;
$^d$ Position of the optical source;
$^e$ Luminosity in the 0.5-4.5 keV band;
\label{addIDs}
\end{sidewaystable}

\section{Discussion}

\subsection{Astrometric accuracy in the X-ray source list} \label{astrometry}

\begin{figure}
\includegraphics[angle=-90,width=8cm]{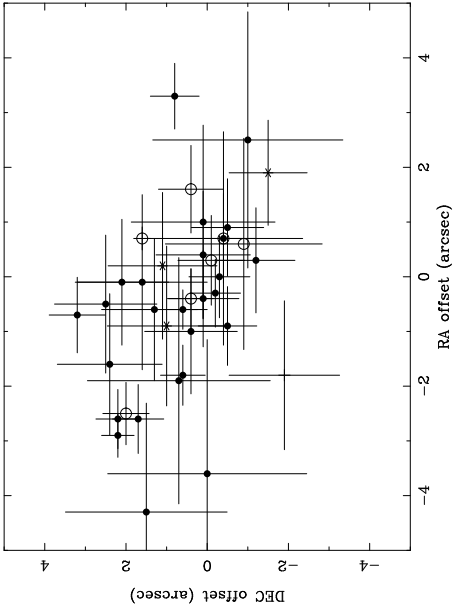}
\caption{Offsets between Optical and X-ray positions. The extended X-ray source \object{XMMU J122017.9+752212} is not shown. Symbols as in \ref{HRs}.}
\label{offsets}
\end{figure}

With the sample of X-ray sources whose optical counterparts have been
positively identified (including the ones below our flux limit),
we check how good the astrometry of the XMM-Newton data was.  We then
analyzed the offsets of the optical position with respect to the X-ray
position. These are shown in Fig. \ref{offsets}, where error bars
are just the X-ray positional errors, which by far dominate over the
optical position ones.

An obvious feature to note is that, apart from the source \object{XMMU J122017.9+752212}
which is extended both in X-rays and in the optical (\object{NGC 4291}), there
is no obvious trend of offsets being larger for any source class.  In
particular the sources classified as NELGs (which in fact are all
AGNs) are unlikely to be chance coincidences, as in that case they
would exhibit larger offsets with respect to the X-ray sources.

\begin{figure}
\includegraphics[angle=-90,width=8cm]{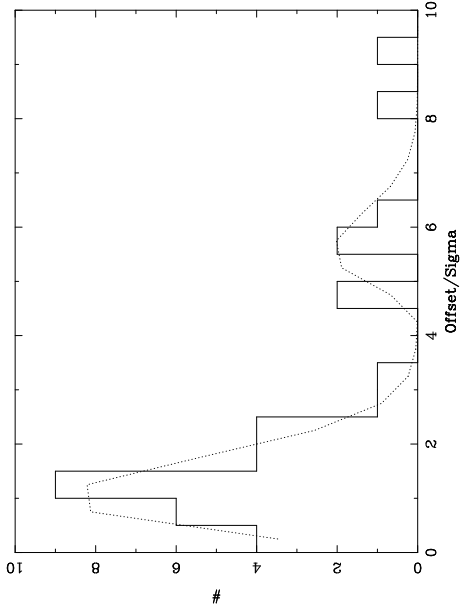}
\caption{Histogram of Offsets (Optical -- X-ray) scaled to the corresponding statistical uncertainty in the X-ray position. The best fit accounting for systematics is also shown}
\label{histo-offsets}
\end{figure}

Fig. \ref{histo-offsets} shows the histogram of these offsets
normalized to their corresponding statistical errors. Assuming that position
errors are well reproduced by a gaussian, this normalized offset $x$
should be distributed as 

\begin{equation}
f_{pos}(x|x_0)\propto \mid x-x_0\mid \exp \left(-\mid
x-x_0\mid^2/2\right)
\end{equation}
where in the absence of systematic shifts $x_0=0$. Actually $x_0$
measures the magnitude of systematic residuals in units of statistical
errors.  An attempt to fit the histogram of fig. \ref{histo-offsets}
without systematics reveals that it can only reproduce the first peak
(at around $x\sim 1$), but leaves the second peak unexplained.  Then
we maximum-likelihood fitted the sum of two of the above functions,
one with $x_0=0$ and a second one with the systematic shift $x_0$ as a
free parameter, the relative contribution of both terms being also
a free parameter.  The result (fit shown in fig. \ref{histo-offsets})
is that about 83\% of the sources have no significant systematic
shifts and the remaining 17\% of the sources have systematic shifts of
$\sim 4.5$ statistical errors (typically $\sim 3-4$ arcsec).

A closer inspection of the sources showing residual astrometric
systematics shows that all of them are in the \object{Mkn 205} field, all of
them are relatively bright and only one of them is relatively close to
the centre of the field (but then close enough to the \object{Mkn 205}
point-spread function tail). We do not fully understand the reason for this, but perhaps
it might be related to the overall astrometric solution being
dominated by some spurious matches of X-ray sources fainter than the
ones used in our survey.  The residuals are, however, so small that
they are irrelevant for a high galactic latitude field.  Much more
care should be taken, however, in galactic plane fields where the
density of optical sources is much larger and chance associations
might produce a completely wrong astrometric solution. We suggest in
this case to use a very limited number of secure identifications to
derive a first astrometric correction and then iterate as the number of
identified sources grows.

\subsection{Overall source populations}\label{global}

\begin{figure}
\resizebox{\hsize}{!}{\includegraphics{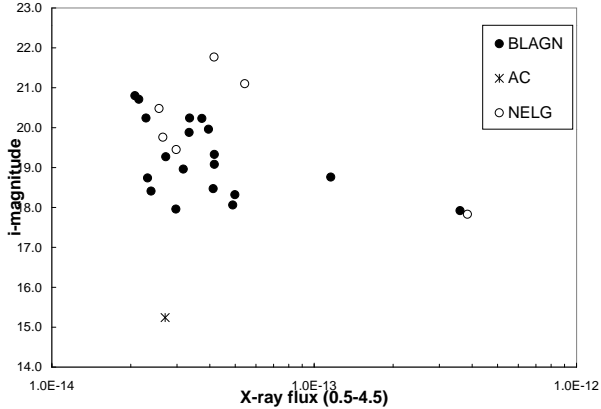}}
\caption{i'-magnitude vs X-ray flux for this sample. Symbols as in \ref{HRs}}
\label{imagvsX}
\end{figure}

As expected, the majority of the identified X-ray sources are
extragalactic (only 1 galactic AC), and in particular AGN.  Fig.
\ref{imagvsX} shows the optical to X-ray relation for the X-ray
sources. We see that most, if not all, of the BLAGNs will be
identified in the medium sensitivity survey (i.e. at a flux limit of
$2\times 10^{-14}\, {\rm erg}\, {\rm cm}^{-2}\, {\rm s}^{-1}$) down to
i'$\sim 22^{\rm m}$.  NELGs and other galaxies are typically
fainter, but we expect to identify large fractions of them within the
AXIS resources. We can conclude from this pilot study that $>90\%$ of
the medium sample can be identified with the AXIS programme.

One remarkable feature of this survey is that we have found no
clusters of galaxies as counterparts of the identified X-ray
sources. Using the source counts from the $ROSAT$ Deep Cluster Survey
\citep{Rosati1998}
we would have expected $<2-3$ clusters in our sample.
We must further stress that there are no cluster candidates in the
unidentified sources, down to the sensitivity limits of the optical
images. Actually, the sliding box
algorithm used to search for sources in the XMM-Newton images is not
optimized for the detection of  extended sources. Therefore we do not expect to be
complete in groups and clusters. In fact, extended sources in
XMM-Newton images are the subject of on-going parallel studies within
the XMM-Newton Survey Science Centre activities.

We have cross-correlated the X-ray source positions with the NVSS
radio survey (Condon et al. 1998).  No coincidences closer than 10
arcsec are found for the flux-limited sample. The only radio source
within 30 arcsec of an X-ray source is a faint source of 2.4 mJy, just
below the completeness level of NVSS, 14.4 arcsec away from \object{XMMU
J010400.9-064949}.  The NVSS positional uncertainty for a source this
faint (rms about 7 arcsec), when combined with the astrometric
accuracy of our X-ray sources (Sect. \ref{astrometry}) make it
unlikely that this is a true association.  For the full sample of 89
sources found by our source-searching algorithms, the number of
coincidences within 30 arcsec increases from zero to one, with a
source of 2.7 mJy being found 7.1 arcsec from an unidentified X-ray
source of 5.6 $\times 10^{-15}\, {\rm erg}\, {\rm cm}^{-2}\, {\rm
s}^{-1}$. Despite the fact that this is likely to be a true
association, preliminary statistics based on about 3000 X-ray sources
from the ChaMP survey (Wilkes et al. 2001; Wilkes and Green 2001,
private communication) find a radio association with NVSS which
increases with X-ray flux, and with which our two XMM-Newton fields
disagree at the 95\% confidence level.  Work is underway to
investigate the radio properties of larger samples of sources.

\subsection{The extragalactic X-ray sources}

\begin{figure}
\resizebox{\hsize}{!}{\includegraphics{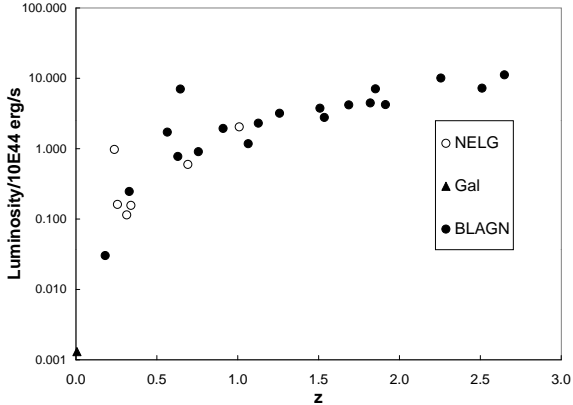}}
\caption{Luminosity-redshift relation for the extragalactic sources
identified in the sample. Symbols as in \ref{HRs}}
\label{Lz}
\end{figure}

Fig. \ref{Lz} shows the luminosity-redshift relation for the
extragalactic sources identified in our sample.  With the exception of
\object{NGC 4291}, which has a luminosity not far from that of a normal galaxy ($\sim
3\times 10^{40}\, {\rm erg}\, {\rm s}^{-1}$), all the remaining
extragalactic objects have luminosities higher than $\sim 10^{42.5}\,
{\rm erg}\, {\rm s}^{-1}$, so there is no doubt they host an Active
Galactic Nucleus.  In particular, all the objects classified as NELGs,
should in reality be called Narrow-line AGNs (NLAGNs). Indeed some (if
not all of them) exhibit [OIII] lines much stronger than [OII] and/or
[NeV] emission lines, all of which are features of a hard non-stellar
ionizing continuum in the narrow-line emitting region.

The weighted average hardness ratios for BLAGN are $\langle
HR_1\rangle=-0.68 \pm 0.01$ and $\langle HR_2\rangle=-0.31 \pm
0.04$. It is interesting to note that for a single power-law spectrum,
these values correspond to $\Gamma\sim 2$ and $\Gamma\sim 1.6$
respectively, i.e., it appears that BLAGN have a hardening in their
spectrum towards high photon energies. 

It is remarkable, however, that the average spectra of BLAGN do not
appear to be consistent with a single power law spectrum.  Indeed,
while $HR_1$ is consistent with an unabsorbed $\Gamma=2$ spectrum, at
harder photon energies $HR_2$ calls for a much flatter X-ray spectrum,
with $\Gamma\sim 1.6$. A similar result has been recently found by
\citet{Pappa2001} by co-adding $ROSAT$ and $ASCA$ spectra of 21
BLAGN. Our result confirms that this is not due to a mismatch between
$ROSAT$ and $ASCA$ calibrations and amphasizes the power of XMM-Newton
in broad-band X-ray spectral studies.

\begin{figure}
\resizebox{\hsize}{!}{\includegraphics{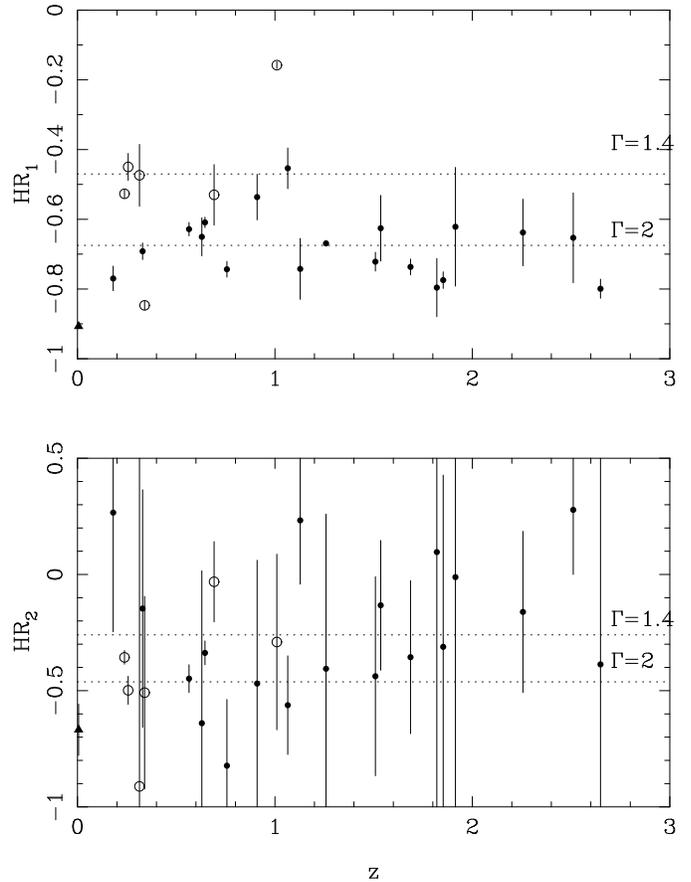}}
\caption{Hardness ratios $HR_1$ and $HR_2$ as a function of redshift
z. Symbols as in \ref{HRs}. We also plot expected values for single
power law spectra with $\Gamma=1.4$ and $\Gamma=2$.}
\label{HRsvsz}
\end{figure}

The spectral shape that we infer for BLAGN was first empirically
proposed by \citet{Schwartz1988} in order to reproduce the XRB
spectrum. The subsequent discovery of a Compton reflection bump in the
X-ray spectrum of Seyfert 1 galaxies \citep{Pounds1990} suggested that
reflection-dominated AGN (i.e., AGN where $>90\%$ of the observed
X-rays would have been reflected in cold/warm material) could
contribute the bulk of the X-ray background
\citep{Fabian1990}. Although no significant population of such sources
has been detected, it is likely that we are witnessing the effects of
Compton reflection or a similar phenomenon in the average X-ray
spectra of the $z\sim 1-2$ BLAGN in our sample. If the Compton
reflection bump entering the 4.5-10 keV band is the ultimate reason
for the hard values of $HR_2$, then a trend with redshift should be
seen. In fig. \ref{HRsvsz} we plot both hardness ratios $HR_1$ and
$HR_2$ as a function of redshift for the extragalactic objects.  No
trend is found in $HR_1$ vs z for BLAGN, and at best we can only claim
a hint of $HR_2$ becoming higher at high redshifts. That, which would
be expected if Compton reflection is responsible for the hardening of
the spectrum, needs much more data to be confirmed.

We only find marginal evidence for the NLAGN having a harder spectrum
than BLAGN (see fig. \ref{HRs}). Such a trend had been found in
$ROSAT$ surveys \citep{Romero1996,Almaini1996}. The weighted average
hardness ratios for these objects are $\langle HR_1\rangle=-0.48 \pm
0.11$ and $\langle HR_2\rangle=-0.37 \pm 0.04$. $\langle HR_2\rangle$
agrees with the corresponding value for BLAGNs, but $HR_1$ is slightly
higher, although only at $\sim 2\sigma$ significance. If we interpret
the marginal difference in terms of a $\Gamma=2$ power-law spectrum
being affected by intrinsic absorption in NLAGN, the inferred column
density is $N_{HI}=5\times 10^{22}\, {\rm cm}^{-2}$ at a redshift of
$z=0.4$ where most of the NLAGNs are located.

We must recall that evidence has been accumulated that X-ray
photoelectric absorption and optical spectroscopic classification do
not appear to have a one-to-one relation. For example hard X-ray
sources found in $ROSAT$ surveys contain large fractions of unobscured
type 1 QSOs and Seyferts \citep{Page2001}. On the contrary, dusty
warm absorbers produce little effect in the X-ray broad-band colours
of AGN, but substantially influence their optical appearence.  The
underlying reason for all these apparent inconsistencies could be the
different distributions of atomic gas and dust in the close
environment of the AGN central engine \citep{Maiolino2001}, perhaps
due to dust sublimation near the center \citep{Granato1997}.

Source \object{XMMU J122017.9+752212} (\object{NGC 4291}) deserves
further comment. X-ray emission from this galaxy was detected with
$Einstein$ \citep{Canizares1987}. \citet{Roberts2000} analyze $ROSAT$
HRI images of nearby galaxies and find a nuclear X-ray point source
within 6.8 arcsec from the centre of \object{NGC 4291}. The extended
X-ray emission of this source in the 0.5-4.5 keV XMM-Newton image
clearly masks out a part of the sky, which is why we have removed a
1.5 arcmin radius circle around it.  However, when this region is
examined only in the 2-10 keV band, the diffuse emission
disappears. What is then seen is a point source approximately
coincident with the nucleus of \object{NGC 4291} and a further 4
sources within the excised region, some of which might possibly be
associated with the galaxy.  The fact that \object{NGC 4291} presents
a point source in its center is a strong suggestion that an active
nucleus might be actually hidden in the centre of this optically dull
galaxy.  A more complete discussion of this interesting galaxy and its
environment is beyond the scope of this paper and will be presented
elsewhere.

\subsection{The Broad-Absorption Line QSOs} \label{BALQSOs}

\begin{figure}
\includegraphics[angle=-90,width=8cm]{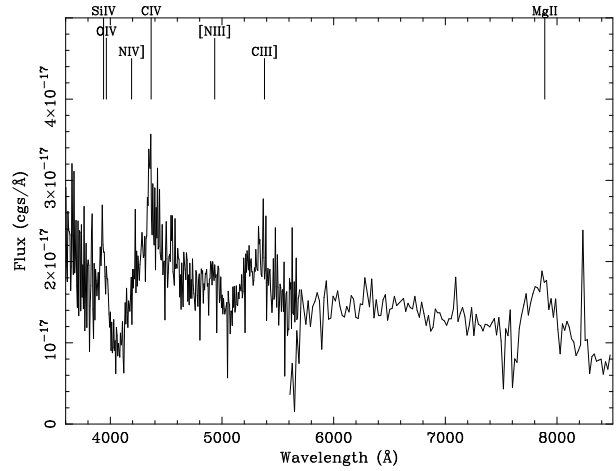}
\caption{Optical spectra of the BAL QSO \object{XMMU J010328.7-064633} obtained with
the ISIS spectrograph on the WHT. Both blue and red arm spectra (which have different channel size) are shown.}
\label{G133pos2_011}
\end{figure}

\begin{figure}
\includegraphics[angle=-90,width=8cm]{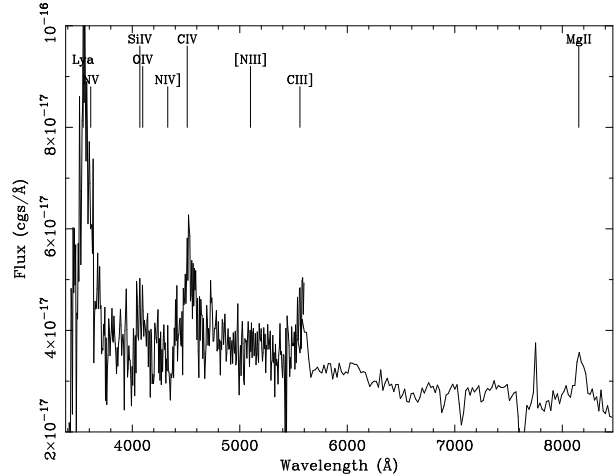}
\caption{Optical spectra of the BAL QSO \object{XMMU J010316.7-065137} obtained with
the ISIS double spectrograph on the WHT. Both blue and red arm spectra (which have different channel size) are shown. }
\label{G133pos2_023}
\end{figure}

Among the sample of BLAGN we find 2 (\object{XMMU J010328.7-064633} at $z=1.82$ and
\object{XMMU J010316.7-065137} at $z=1.91$) Broad-Absorption-Line (BAL) QSO (see figs.
\ref{G133pos2_011} and \ref{G133pos2_023}). Parameters of the
corresponding broad-absorption line systems (based on the CIV line)
are listed in Table \ref{BALparams}. In particular we have computed
the so-called {\it Balnicity index} introduced by \citet{Weymann1991}
as a way to provide a continuous classification between BAL and
non-BAL QSOs. In the context of the sample studied by
\citet{Weymann1991} the value of the Balnicity Index is in the low end for \object{XMMU
J010316.7-065137} but close to average for \object{XMMU J010328.7-064633}.

\begin{table}
\caption{Parameters of the Broad absorption systems found in the BAL
QSOs, based on the CIV line: $W_{rest}$ is the rest-frame equivalent width of the
broad-absorption line, $v_{ej}$ its ejection
velocity from the QSO and BI is the {\it Balnicity index}
defined in \citet{Weymann1991}.}
\begin{tabular}{l c c c}
Source           & $W_{rest}$ & $v_{ej} $ & BI\\
                 &  (\AA ) &  (${\rm km}\, {\rm s}^{-1}$) & (${\rm km}\, {\rm s}^{-1}$)\\  \hline
\object{XMMU J010316.7-065137}    & 5                     & 19000 &  580\\
\object{XMMU J010328.7-064633}    & 34                    & 34000 & 3650\\ \hline
\end{tabular}
\label{BALparams}
\end{table}

BAL QSOs  have been practically absent from
previous X-ray selected samples. There have been a few exceptions: 1 BAL QSO in
the Einstein Medium Sensitivity Survey at $z=2.027$ \citep{Stocke1991}
and more recently the ELAIS/BeppoSAX survey at $z=2.2$
\citep{Alexander2001}, the {\it Chandra} deep field south at $z=2.75$,
\citep{Giacconi2001} and the survey of $ROSAT$ hard X-ray sources
at $z=2.21$ \citep{Page2001}.  

The paucity of BAL QSOs in previous X-ray surveys is hardly surprising, as the gas ejected by BAL QSOs which is
responsible for the broad absorption lines will efficiently absorb soft
X-rays. Previous X-ray missions, mostly sensitive to soft X-rays like
$Einstein$ and $ROSAT$, are expected to have a much reduced sensitivity
to BAL QSOs with respect to optical surveys.  It is illustrative to
notice that the only $ROSAT$-discovered BAL QSO was found in a selection
of hard X-ray spectrum sources. 

We find it intriguing that the BAL QSOs have  `normal' values
of $HR_1$: $-0.79\pm 0.08$ for \object{XMMU J010328.7-064633} and $-0.62\pm 0.17$ for
\object{XMMU J010316.7-065137} (the average over all sources in the sample is $\sim
-0.69\pm 0.01$ for $HR_1$). Indeed, the high redshift of the two
objects helps to move the intrinsic photoelectric absorption feature
below our detection band (0.5-4.5 keV). Assuming $z=1.85$, and an
unabsobed $\Gamma=2$ power-law spectrum, an intrinsic absorbing column
of $\sim 10^{22}\, {\rm cm}^{-2}$ increases the value of $HR_1$ from -0.68 in the
absence of intrinsic absorption to -0.61. It is then clear that these
objects cannot have intrinsic columns significantly in excess of
$10^{22}\, {\rm cm}^{-2}$ or otherwise we would see it in their X-ray spectra.  
However, these are unusually small values for the absorbing columns in
BAL QSOs.  \citet{Gallagher2001} report on ASCA and $Chandra$
observations of several BAL QSOs, and the inferred absorbing columns
always exceed $10^{22}\, {\rm cm}^{-2}$ and sometimes by a large
amount. Here we see that low (neutral) column density BAL QSOs do
exist. Perhaps high ionisation and/or partial covering could bring to agreement
the apparent small X-ray absorbing column with a sizeable
CIV broad absorption line.

A further remarkable fact is that both BAL QSOs lie at the redshift
where the distribution of BLAGN peaks. This might be telling us that we are
only seeing the tip of the iceberg, i.e., we have only detected BAL QSOs
at the redshift where this detection would be more likely (helped
indeed by the K-correction discussed above).

What remains to be understood is why BAL QSOs were practically absent
in $ROSAT$ surveys of similar depth (e.g. Boyle et al 1994).  The
sensitivity to higher X-ray photons of our current survey (0.5-4.5 keV)
with respect to the standard 0.5-2 keV $ROSAT$ band does not make a
big difference: for a $z=1.85$, intrinsically absorbed ($10^{22}\,
{\rm cm}^{-2}$) power-law ($\Gamma=2$) QSO spectrum, only $\sim 20\%$ of
the counts from the source fall in the 2-4.5 keV band in EPIC pn. A source
with that spectrum at our survey limit, would have a 0.5-2 keV flux
slightly above $\sim 1\times 10^{-14}\, {\rm erg}\, {\rm cm}^{-2}\,
{\rm s}^{-1}$ and would have been therefore detectable in the
so-called $ROSAT$ deep survey (Boyle et al 1994).


\section{Conclusions}

As expected, the majority (25 out of 29 or 86\%) of the X-ray sources
identified in this sample are AGN. These are divided into 25\%
Narrow-line AGN and 75\% Broad-Line AGN. Narrow-line AGNs have some
hints of photoelectric absorption with respect to the BLAGN
population, but the effect is less than $\sim 2\sigma$ significant in
our small sample. Inferred columns for narrow-line AGNs at $z\sim 0.4$
would be around $\sim 5\times 10^{22}\, {\rm cm}^{-2}$.

Perhaps the most remarkable finding is that 2 out of 24 AGN are BAL
QSOs, i.e., $\sim 10\%$.  This is similar to the fraction of BAL QSOs
in optically selected samples. The X-ray spectrum of these BAL QSOs
does not show evidence for photoelectric absorption, and the limits we
derive for the intrinsic absorption column ($< 10^{22}\, {\rm
cm}^{-2}$) tell us that these are unusual objects among the BAL QSO
population in their X-ray spectra.

The AXIS high-galactic latitude medium survey is now progressing with
$\sim 250$ X-ray sources spectroscopically identified. Large
numbers of sources are needed, not only to study in detail the overall
properties of the dominant source population (AGN in this case), but
also to find and characterize the rarer populations, e.g., BAL QSOs or
optically dull galaxies \citep{Mushotzky2000,Baldi2001}.

\begin{acknowledgements}
We thank Alberto Fern\'andez-Soto for help with the photometric
redshift techniques and the referee Ioannis Georgantopoluos for
interesting suggestions. We are grateful to the CCI of the observatories
of the Canary Islands for a generous allocation of telescope time,
through the International Time Programme scheme. The INT/WHT and TNG
telescopes are operated on the island of La Palma by the Isaac Newton
Group and the Centro Galileo Galilei respectively in the Spanish
Observatorio del Roque de Los Muchachos of the Instituto de Astrof\'\i
sica de Canarias. XB, FJC, MTC and SM acknowledge financial support by the
Spanish MCYT under project AYA2000-1690. RDC and TM acknowledge
partial financial support by the Italian Space Agency (ASI) and by the
MURST (Cofin00-32-36).  This project was supported by the DLR under
grants 50 OR 9908 0 (GPS) and 50 OX 9801 3 (HB).
\end{acknowledgements}

\bibliographystyle{apj}

\end{document}